\documentclass[twoside,twocolumn,english,superscriptaddres,showpacs,longbibliography,aps,prb]{revtex4-2}

\usepackage[T1]{fontenc}
\usepackage[utf8]{inputenc}
\usepackage[english]{babel}
\usepackage{bm}
\usepackage{amsmath}
\usepackage{amssymb}
\usepackage{graphicx}
\usepackage[unicode=true, pdfusetitle, bookmarks=true,bookmarksnumbered=true,bookmarksopen=true,bookmarksopenlevel=1,
breaklinks=false,pdfborder={0 0 0},pdfborderstyle={},backref=false,color links, citecolor=blue,linkcolor=red,urlcolor=blue] {hyperref}
\usepackage[nameinlink]{cleveref}
\Crefname{figure}{Fig.}{}

\makeatletter
\def\set@firstnote#1{%
 \@ifnum{\firstnote@num=#1\relax}{}{%
  \class@warn@end{Endnote numbers changed: rerun LaTeX}%
 }%
 \immediate\write\@mainaux{%
   \global\mathchardef\string\firstnote@num#1\relax
 }%
}%

\usepackage{braket}
\usepackage{bm}
\usepackage{tikz}
\usepackage{pgfplots}
\usepackage{pgf}
\usepackage{subfigure}
\usepackage{nicematrix}
\usepackage{diagbox}
\usepackage{orcidlink}

\renewcommand{\selectlanguage}[1]{}

\makeatother

\pgfplotsset{compat=1.18}

\begin{document}

\title{Fusion rule in conformal field theories and topological orders: A unified view of correspondence and (fractional) supersymmetry and their relation to topological holography}
\author{Yoshiki Fukusumi}
\affiliation{Center for Theory and Computation, National Tsing Hua University, Hsinchu 300044, Taiwan}
\affiliation{Physics Division, National Center for Theoretical Sciences, National Taiwan University, Taipei 106319, Taiwan}

\date{\today}
\begin{abstract}
The algebraic or ring structure of anyons, called the fusion rule, is one of the most fundamental research interests in contemporary studies on topological orders (TOs) and the corresponding conformal field theories (CFTs). Recently, the algebraic structure realized as generalized symmetry, including non-invertible and categorical symmetry, captured attention in the fields. Such non-abelian anyonic objects appear in a bulk CFT or chiral CFT (CCFT), but it has been known that the construction of a CCFT contains theoretical difficulties in general. In this work, we propose the fusion rules in $Z_{N}$ extended chiral and bulk conformal field theories and the corresponding TOs. We explicitly construct a nontrivial expression of subalgebra structure in the fusion rule of a bulk CFT. We name this subalgebra ``bulk semion". This corresponds to the fusion rule of the CCFT and categorical symmetry of the TO or $Z_{N}$ graded symmetry topological field theory (SymTFT). This gives a bulk-edge correspondence based on the symmetry analysis and corresponds to an anyon algebraic expression of topological holography in the recent literature. The recent topological holography is expected to apply to systems in general space-time dimensions. Moreover, we present a concise way of unifying duality (or fractional supersymmetry), generalized or categorical symmetry, and Lagrangian subalgebra. Our method is potentially useful to formulate and study general TOs, fundamentally only from the data of bulk CFTs or vice versa, and gives a clue in understanding CCFT (or ancillary CFT more generally).
\end{abstract}

\maketitle

\section{introduction}

Non-abelian anyons and their (fusion) algebra are one of the central research interests in contemporary physics, including condensed matter and high-energy physics and quantum information\cite{Kitaev:1997wr,nayak_non-abelian_2008}. This structure plays a fundamental role in describing the structure of excitations, (generalized) symmetries, and the corresponding operators in topological orders (TOs) or topological quantum field theories (TQFTs) and the underlying conformal field theories (CFTs)\cite{Witten:1988hf,Moore:1991ks}. Such objects appear ubiquitously in contemporary physics and mathematics, so we only note earlier literature on the analysis of correlation functions or partition functions \cite{Verlinde:1988sn,Cardy:1989ir,Fuchs:1993et}and the topological defects\cite{article,Bockenhauer:1999ae,Bockenhauer:1998ef,Bockenhauer:1999wt,Petkova:2000ip} in CFTs. We also note the recent review articles and lecture notes\cite{McGreevy:2022oyu,Cordova:2022ruw,Schafer-Nameki:2023jdn,Bhardwaj:2023kri,Shao:2023gho,Carqueville:2023jhb}.

Recently, categorical symmetry or symmetry topological field theory (SymTFT) \cite{Ji:2019jhk,Apruzzi:2021nmk,Chatterjee:2022tyg,Bullimore:2024khm} has captured attention as a symmetry-based description of TOs. For example, the double semion algebra in the $1+1$ dimensional Majorana-Ising CFT or in the $2+1$ dimensional $Z_{2}$ TO has been studied. Historically, the correspondence between the double semion algebra and the Majorana-Ising CFT appeared in the study of fermionic string theories\cite{Ginsparg:1988ui}. However, generalization to a $Z_{N}$ symmetric model has not been studied except for several works\cite{Schoutens:2015uia,Lan2016ModularEO,Fukusumi:2022xxe,Fukusumi:2023psx}, regardless of the fundamental importance for the understanding of properties of anyons. It is worth stressing that the resultant $Z_{N}$ extended chiral CFT (or fractional supersymmetric string theory in the context of $Z_{N}$ nonanomalous models as in \cite{Gliozzi:1976qd}) is outside of existing modular tensor categories (MTCs) and has been lacking general mathematical frameworks\cite{Lan2016ModularEO,Galindo:2024qzg,Gannon:2024tcl}. In other words, surprisingly, even the fusion algebraic data of anyons corresponding to fractional quantum Hall (FQH) states, $Z_{N}$ graded SymTFTs, have not been studied sufficiently (Related unusual properties have been clarified more recently in \cite{Fukusumi:2023psx}). It is also worth noting that an FQH state is regarded as a typical example of symmetry-enriched topological orders in more recent literature. More phenomenologically,the chiral correlation functions of the anomaly-free $Z_{N}$ particle, called (half) integer spin simple current appearing in the studies on discrete torsion and extended models\cite{Vafa:1986wx,Hamidi:1986vh,Schellekens:1989am,Schellekens:1990ys,Gato-Rivera:1990lxi,Gato-Rivera:1991bqv,Gato-Rivera:1991bcq}, define $Z_{N}$ generalization of the wavefunctions of the corresponding FQH states or TQFTs through the bulk-edge correspondence or CFT/TQFT correspondence\cite{Laughlin:1983fy,Witten:1988hf}. The corresponding fusion rules are nothing but $Z_{N}$ graded SymTFTs in a modern technical term, and this unusual structure outside of bosonic theory has been studied several times, for example in \cite{Lu:2010vda,Cappelli:2010jv,Schoutens:2015uia}.

In this work, we propose a new $Z_{N}$ extended version of general correspondence between a bulk CFT (product of chiral and antichiral CFTs) or spherical fusion category (SFC) and a chiral CFT (CCFT) or TO (or SymTFT) at the level of the fusion rule of anyonic objects. Starting from a $Z_{N}$ symmetric bulk CFT and its fusion algebra, we show a nontrivial construction of subalgebra, $Z_{N}$ graded SymTFT denoted as $\mathbf{S}$, based on its symmetry charge (or quantum anomaly) and parity (or particle counting) of the objects. We name this subalgebra ``bulk semion algebra," corresponding to the algebraic (or categorical) structure of the TOs. The CFT/TQFT results in the correspondence between the bulk semion algebra and the fusion rule of CCFT. Our formalism provides the necessary algebraic data formulating ``topological holography" \cite{Ishtiaque:2018str,Moradi:2022lqp,Bhardwaj:2023bbf,Huang:2023pyk,Inamura:2023ldn,Bhardwaj:2024qrf,Wen:2024udn}, symmetry-based representation of the CFT/TQFT \cite{moore_nonabelions_1991,Witten:1988hf,Fuchs:2002cm,Fuchs:2003id,Fuchs:2004dz,Chen:2022wvy} (However, we note that the term ``topological holography" appeared earlier to notify the system both topological and holographic in \cite{Husain:1998vz,deMedeiros:2001wqm,Li:2019qzx,li:tel-03269600} in a different context. Similar terminological issues have been summarized in the introduction of \cite{Chatterjee:2024ych}. We also note that the almost same term has been used in optics  with different meanings\cite{Yu_2021,kong2023topologicalholographystorageoptical}). Moreover, our formalism is much more concise and general, with intuitive and rigorous algebraic expressions with a concrete identification with modular partition functions, Eq.\eqref{modular_covariant}, and analysis on $Z_{N}$ anomaly or \emph{fractional supercharge} defined by Eq.\eqref{fractional_supercharge} that are absent in references in category theory. The corresponding category theories are still under development, and our explicit method from the established techniques in abstract algebra (or linear algebra) will provide a clue in constructing such extended category theories in a coherent way.

The main proposal of the present manuscript can be summarized as the following schematic picture,
\begin{equation}
\begin{split}
&\{ Z_{N} \text{ symmetric SFC}\} \\
&\rightarrow \{ Z_{N} \ \text{extended SFC}\} \Rightarrow \{ Z_{N} \  \text{graded SymTFT}\} 
\end{split}
\end{equation}
where $\rightarrow$ represents the $Z_{N}$ extension and $\Rightarrow$ represents the bulk semionization. The arrow $\rightarrow$ transforms a $Z_{N}$ symmetric bosonic bulk CFT to a $Z_{N}$ extended bulk CFT analogous to a quark-hadron-like model, or a fractional supersymmetric model. This operation itself has been studied in category theory\cite{turaev2000homotopyfieldtheorydimension,etingof2009weaklygrouptheoreticalsolvablefusion}, but the resultant partition functions, Eq. \eqref{modular_covariant}, and their anomaly-free conditions have not been studied in this context to our knowledge. In abstract algebra, the topological holography is just a combination of the group extension (or the operation taking a tensor product between a group ring and SFC) and the operation taking a subalgebra in our settings, and this can be expressed as the combination of the two arrows. This simplification by abstract algebra provides a powerful theoretical framework for exact computations of fusion coefficients in the unexplored topological orders, $Z_{N}$ graded SymTFTs. We also note that this kind of reduction from a bulk CFT($\sim \text{CCFT} \times \overline{\text{CCFT}}$) to a CCFT can be interpreted as a massive renormalization group flow\cite{Date:1987zz,Saleur:1988zx,Foda:2017vog} which has a close connection to the Li-Haldane conjecture in condensed matter physics\cite{Li_2008,Qi_2012,Cardy:2017ufe} and Moore-Seiberg data in high energy physics\cite{Moore:1988qv,Moore:1989vd}. In $2+1$ dimensional TOs or TQFTs, the above arrow $\Rightarrow$ corresponds to the gluing process of chiral and antichiral edges (See also the corresponding discussions in \cite{Wen:1990se,Qi_2012,Fukusumi:2022xxe}). We also remind that the TOs with two edges, or those in the cylinder geometry, correspond to modular partition functions. Historically, the significance of the cylinder geometry and its correspondence to the bulk CFT have been established in earlier literature\cite{Milovanovic:1996nj,Cappelli:1996np}.

The rest of the manuscript is organized as follows. Sec. \ref{semionization} is the main part of this work, and we propose a construction of the bulk semion algebra or $Z_{N}$ graded SymTFT $\mathbf{S}$ which corresponds to the fusion rule of anyonic objects in $2+1$ dimensional $Z_{N}$ TOs along with discussions on the less familiar (or new) modular partition functions, Eq.\eqref{modular_covariant}. In Sec.\ref{TO-CCFT}, we discuss the application of our method to construct correspondence between TOs and the CCFTs. Consistency between our method and existing chiral algebra is checked in the Majorana CFT. Moreover, we report a new representation of the bulk CFT which can be related to the (fractional) supersymmetry or related even-odd problem on the lattice model along with the unusual degenerate fusion rule. In Sec. \ref{relation_algebra_category}, we show the precise relationship between our algebraic formalism and existing (and conjectural) category theories. This section provides an overview of the understanding of TOs with respect to anyon algebra. In Sec.\ref{conclusion}, we make a concluding remark by showing a figure of the correspondence between CFTs and TOs, so-called topological holography, potentially applicable to systems in general space-time dimensions. In Appendix \ref{new_example}, we apply the method in the main text to the $SU(3)_{3}$ Wess-Zumino-Witten model, which is less familiar, and demonstrate a new type of fusion rules.

\section{Bulk semionization}
\label{semionization}
In this section, we present the main results of this work, the $Z_{N}$ symmetric bulk semionization of anomaly-free $Z_{N}$ two-dimensional conformal field theories (summarized as the construction of a SymTFT $\mathbf{S}^{(l)}$ from an SFC $\mathbf{F}^{(l)}$). This provides the algebraic way to relate topological symmetry or SFC and categorical symmetry or SymTFT, enabling one to calculate the fusion coefficients in general. More specifically, the derivation of $\mathbf{S}^{(l)}$ from $\mathbf{F}^{(l)}$ is new. Related approaches can be seen in \cite{Huang:2023pyk,Wen:2024udn,Bhardwaj:2024ydc,Huang:2024ror}, but the fusion coefficients of the $Z_{N}$ graded SymTFTs are impossible (or very difficult at least) to obtain straightforwardly from their methods. We also note that the corresponding extended vertex operator algebra has been studied in the studies of monstrous moonshine\cite{Borcherds:1983sq}, but their method is limited to some specific central charges, such as $c=16$ or $c=24$ (see an earlier work \cite{Dong_1996} and recent works \cite{Galindo:2024qzg,Gannon:2024tcl}, for example). In the following discussion, we use the same notations as in the author's previous works\cite{Fukusumi_2022_c,Fukusumi:2023psx} and note the reviews \cite{Fuchs:1997af,Schweigert:2000ix} for earlier literature on the simple current extension.

First, let us concentrate our attention on a $Z_{N}$ symmetric (bosonic) CFT with $Z_{N}$ simple current $J$, where the $Z_{N}$ simple current is a chiral primary field acting like a generator of the $Z_{N}$ group. We introduce the $Z_{N}$ charge of a chiral operator $\phi_{\alpha}$ labelled by $\alpha$ as,
\begin{equation}
Q_{J}(\alpha)= h_{\alpha}+h_{J}-h_{J\times \alpha},
\label{fractional_supercharge}
\end{equation}
where $``\times"$ represents the fusion of objects, $h$ is the conformal dimension, and we do not distinguish the label and the operator itself to simplify the notations. For example, in a $Z_{2}$ model, the charge zero sector corresponds to the Neveu-Schwartz sector and the charge $1/2$ sector corresponds to the Ramond sector. Hence, the quantity $Q_{J}(\alpha)$ naturally define $Z_{N}$ analog of supercharge, fractional supercharge, or \emph{$Z_{N}$ anomaly}.

The anomaly-freeness of this $Z_{N}$ theory can be defined as the condition $Q_{J}(J^{n})=0$
for all $n=0,1,...,N-1$ where $J^{n}$ is the $n$ times multiplication of $J$. In other words, the $Z_{N}$ symmetry itself is not charged (or not mutually twisted) in the anomaly-free theory, and one can identify the anomaly-free theory as a fractional supersymmetric model by generalizing the arguments on the $Z_{2}$ cases. This condition can be satisfied under the following conditions,
\begin{itemize}
\item{$h_{J}=\text{integer}$ for both $N$ even and odd.}
\item{$h_{J}=\text{half-integer}$ for $N$ even.}
\end{itemize}
The models satisfying these conditions have been known as discrete torsion or integer simple current extension of CFTs  \cite{Vafa:1986wx,Hamidi:1986vh,Schellekens:1989am,Schellekens:1990ys,Gato-Rivera:1990lxi,Gato-Rivera:1991bqv,Gato-Rivera:1991bcq}. For example, this condition can be realized in the $SU(N)_{N}$ Wess-Zumino-Witten (WZW) model\cite{Wess:1971yu,Witten:1983tw}. More recently, its relation to $SU(N)$ Haldane conjecture\cite{Haldane:1983ru,Haldane:1982rj,Wamer:2019oge} or Lieb-Schultz-Mattis theorem has been revisited in \cite{Kikuchi:2022ipr,Fukusumi_2022_c}. For the readers interested in the anomaly analysis of the $Z_{N}$ symmetric models and their renormalization group (RG) flows, we note several recent works \cite{Furuya:2015coa,Lecheminant:2015iga,Cho:2017fgz,Numasawa:2017crf,Yao:2018kel,Alavirad:2019iea,Kikuchi:2019ytf,Li:2022drc,Cordova:2022lms,Gannon:2023udg,Lecheminant:2024apm} and earlier related works\cite{PhysRevB.46.2896,Alcaraz_1989,PhysRevB.34.6372,Schulz,Cabra:1998vw} and note recent works on noninvertible symmetry \cite{Kikuchi:2023gpj,Kikuchi:2023cgg,Jacobsen:2023isq,Kikuchi:2024hwf,Kikuchi:2024pex,Cordova:2024vsq,Bhardwaj:2024kvy,Bhardwaj:2024wlr,Chatterjee:2024ych}. The anomaly-free $Z_{N}$ models play a central role in constructing $Z_{N}$ TOs or FQH states \cite{Blok:1991zq,Cappelli:2010jv,Schoutens:2015uia,Dorey:2016mxm,Fuji_2017,Fukusumi_2022_c,Henderson:2023kjz,Bourgine:2024ycr} with close analogy and connection to quark-hadron models or \emph{fractional supersymmetric models} \footnote{Following the terminologies to call fermionic (or $Z_{2}$ extended) fusion category as superfusion category, we call $Z_{N}$ extended model as fractional supersymmetric model. We also note that the $Z_{N}$ anomaly-freeness in this manuscript can be regarded as a natural extension of the notion of the anomaly-freeness in fermionic (or supersymmetric) models.}. In the contexts of the FQH states and TOs, the integer simple currents define the $Z_{N}$ analog of electron operators defining the wavefunctions, and this aspect is the most fundamental in constructing the wavefunctions\cite{Laughlin:1983fy}.

We assume the sectors of the bosonic $Z_{N}$ CFT are decomposed as $Z_{N}$ noninvariant sector specified as $ i,p$, and $Z_{N}$ invariant sectors $a$, where $p$ denotes $Z_{N}$ parity for simplicity. For the $Z_{N}$ noninvariant chiral object $\phi_{i,p}$, the relation $J\times\phi_{i,p}=\phi_{i,p+1}$ holds. For the $Z_{N}$ invariant chiral object $\theta_{a,0}$, the relation $J\times\theta_{a,0}=\theta_{a,0}$ holds (For a notational reason, we introduced the label $0$ for $\theta$). The decomposition of the sectors is valid for the case when $N$ is a prime number, for example, and the method in the present paper can be applied for more general cases by considering the subgroup structure of $Z_{N}$ \footnote{More detailed analysis will be presented in future work by the author}. The bosonic charge conjugated modular invariant partition function on a torus is,
\begin{equation} 
Z= \sum_{i, p}\chi_{i,p}(\tau) \overline{\chi}_{\overline{i}, -p}(\overline{\tau})+ \sum_{a}\chi_{a}(\tau) \overline{\chi}_{\overline{a}}(\overline{\tau}),
\end{equation}
where $\chi$ is the chiral character labelled by the indices $i,p$ and $a$ and $\tau$ and $\overline{\tau}$ are the modular parameters.

Corresponding to the bosonic modular invariant, the following set of objects, $\mathbf{B}$, forms SFC\cite{Petkova:2000ip}, 

\begin{align}
\Phi_{i,p,-p}&. \\
\Phi_{a,0}&.
\end{align}
The algebraic data of this category theory are called nonchiral fusion rules in older literature \cite{Rida:1999ru,Rida:1999xu}, and this has a close connection to the conformal bootstrap\cite{Polyakov:1974gs}. To avoid confusion, we have avoided the chiral-antichiral decomposition of the sectors. In literature, the corresponding decomposition is called the Deligne product, and it should be distinguished from the tensor product. We note an older paper \cite{Fuchs:1993et} explaining the corresponding structure and recent works\cite{Kong:2019byq,Huston:2022utd}. We also note \cite{Nivesvivat:2025odb} as a recent analysis on the nonchiral fusion rule and conformal bootstrap. In the present manuscript, the property of associativity plays no role evidently, and it seems possible to apply our method to the nonassociative fusion rule in  \cite{Nivesvivat:2025odb} in principle. The fusion rule without associativity (or nonassociative algebra) is called ``magma" \cite{bourbaki1998algebra,Bergman1996CogroupsAC}, and this mathematical structure and its application in physics have not been studied well as far as we know\footnote{We thank Sylvain Ribault and Ingo Runkel for the corresponding discussions.}.  

The ring isomorphism between this SFC and the MTC is widely known as Moore-Seiberg data\cite{Moore:1988qv,Moore:1989vd}. In more recent literature in mathematics, this correspondence has been verified by the arguments based on the boson condensation in \cite{Bais:2008ni,Kong:2013aya,Kawahigashi:2021hds}. For later use, we introduce the label of bosonic CCFT or MTC as $\{ \phi_{i,p} , \theta_{a,0}\}$ and represent the ring isomorphism as,
\begin{equation}
\{ \phi_{i,p} , \theta_{a,0}\}\leftrightarrow \{ \Phi_{i,p,-p}, \Phi_{a,0}\}
\end{equation}
In other words, the algebraic data of the SFC are constrained by the algebraic data of the MTC, and one can consider that a class of SFCs is determined by the combination of the Verlinde formula \cite{Verlinde:1988sn,Cardy:1989ir,Fuchs:1993et} and Moore-Seiberg data \cite{Moore:1988qv,Moore:1989vd}. This can be useful for the readers to understand the SFC from the knowledge of MTC or CCFT. We also note that this ring isomorphism is a fundamental structure introducing the Deligne product as a reduction from tensor products of chiral and antichiral MTCs, $\mathbf{M}\otimes \overline{\mathbf{M}}$, to the SFC $\mathbf{B}=\mathbf{M}\boxtimes\overline{\mathbf{M}}$ where $\mathbf{M}$ represents a MTC. We primarily discuss the tensor product $\otimes$, which realizes the extensions and their algebraic structure, and the fusion product $\times$ in subsequent discussions.

The simple current extended bulk CFT $\mathbf{F}$ \cite{Vafa:1986wx,Hamidi:1986vh,Schellekens:1989am,Schellekens:1990ys,Gato-Rivera:1990lxi,Gato-Rivera:1991bqv,Gato-Rivera:1991bcq} can be obtained by introducing the parity shift operation \cite{Runkel:2020zgg} (or recursive multiplications of $J$ and $\overline{J}$) to the bosonic theory $\mathbf{B}$. Alternatively, the simple current extension of CCFT can be realized by decomposing the $Z_{N}$ invariant sector as $\theta_{a,0}=\sum_{p=0}^{N-1}\phi_{a,p}/\sqrt{N}$ where the extended object $\phi_{a,p}$ is the $Z_{N}$ generalization of chiral semion\cite{Ginsparg:1988ui,Lou:2020gfq,Fukusumi_2022_c,Fukusumi:2023psx}. Because of the introduction of this extended object, the extended CCFT does not correspond to an MTC.
Under the operator-state correspondence\cite{Fukusumi:2023psx}, the bulk operators can be specified as,
\begin{align}
\Phi_{i,p,\overline{p}}&=\phi_{i,p}\overline{\phi}_{\overline{i},\overline{p}} \left( = J^{p}\overline{J}^{\overline{p}}\Phi_{i,0,0} \right), \\
\Phi_{a,p}&=\sum_{p'}\frac{\phi_{a,p'+p}\overline{\phi}_{\overline{a},-p'}}{\sqrt{N}} \left( = J^{p}\Phi_{a,0}=\overline{J}^{p}\Phi_{a.0} \right).
\end{align}
The algebra of $\Phi$ with these extended indices corresponds to the $Z_{N}$ extended SFC\cite{turaev2000homotopyfieldtheorydimension,kirillov2001modularcategoriesorbifoldmodels,mueger2004galoisextensionsbraidedtensor,bischoff2021computingfusionrulesspherical,davydov2021braidedpicardgroupsgraded} and we note the theory as $\mathbf{F}$ which can be interpreted as a $Z_{N}$ particle system or fractional supersymmetric model. In this construction, the addition of $J$ or $\overline{J}$ corresponding to the assignment of the $Z_{N}$ version of the spin structure plays the fundamental role. We stress that, at the algebraic level, the extended theory $\mathbf{F}$ is isomorphic to $Z_{N}\otimes \mathbf{B}$ (or the tensor product of the group ring $Z_{N}$ and $\mathbf{B}$) where $\otimes$ is just the tensor product (not the Deligne product). The group ring structure corresponds to the procedure of stacking symmetry-protected-topological phase in the literature\cite{Kapustin:2017jrc,Karch:2019lnn}. This provides a phenomenological understanding of the extension as a generalization of the Jordan-Wigner or Fradkin-Kadanoff transformations as in \cite{Yao:2020dqx}.

Because the objects $J$ or $\overline{J}$ are outside of the original model $\mathbf{B}$, the extended theory $\mathbf{F}$ cannot be described by SFC, whereas their algebraic data are well-defined. We also remark that the chiral-antichiral tensor products are consistent only up to the characters and partition functions at this stage. We note again that the product between chiral and antichiral fields is interpreted as a kind of Deligne product, not as a tensor product.

For the readers interested in the explicit calculations of the fusion coefficients of the extended model, we note a useful relation\cite{Fukusumi:2025ljx},
\begin{equation}
J\times (\Phi_{\alpha}\times \Phi_{\beta})=\sum_{\gamma}N^{\gamma}_{\alpha, \beta}J\Phi_{\gamma}
\label{useful_relation}
\end{equation}
where the Greek simbols, $\alpha,\beta,$ and $\gamma$ takes arbitrary value $(i,p,\overline{p})$ or $(a,p)$. By choosing $\alpha$ and $\beta$ as original labels in the bosonic theory $\mathbf{B}$, one can easily check the usefulness of the above relation\footnote{We thank Yuma Furuta and Shinichiro Yahagi for the related discussions.}.

By changing the definition of $Z_{N}$ parity for $Z_{N}$ invariant sector (or replacing the order and disorder field), one can relate the $Z_{N}$ Kramers-Wannier (KW) duality in $\mathbf{F}$ to $Z_{N}$ T-duality in $\mathbf{S}$ (for the $Z_{2}$ case, see \cite{Seiberg:2023cdc,Runkel:2020zgg}). For the readers unfamiliar with the arguments, we summarize the notions of bulk CFTs and the partition functions.
\begin{align}
\text{Parity zero sector}& =\text{Bosonic modular invariant},\\
\text{Charge zero sector}& =\text{$Z_{N}$ extended modular invariant},\\
\text{$Z_{N}$ duality}& \sim\text{Changing parity of the sector $a$} 
\end{align}
Especially, the application of the third point on the duality to the models in \cite{Li:2023mmw,Li:2023knf,Yan:2024eqz,Bhardwaj:2024ydc,Huang:2024ror} will be useful because of the conciseness of our formalism. Related discussions can be seen in the author's previous works\cite{Fukusumi_2022_c,Fukusumi:2023psx,Fukusumi:2023vjm}. We also note that, corresponding to the $Z_{N}$ extension, the form of the modular partition function is changed to 
\begin{equation}
Z_{Q}=\sum_{Q_{J}(i)=Q }|\sum_{p}\chi_{i,p}|^{2}+N\sum_{Q_{J}(a)=Q }|\chi_{a}|^{2},
\label{modular_covariant}
\end{equation}
where $Q$ is the $Z_{N}$ charge of the sectors, or fractional supercharge of the model. By restricting our attention to the uncharged sector with $Q=0$, the partition function completely reproduces the existing ones in the literature\cite{Vafa:1986wx,Hamidi:1986vh,Schellekens:1989am,Schellekens:1990ys,Gato-Rivera:1990lxi,Gato-Rivera:1991bqv,Gato-Rivera:1991bcq}. In other words, we introduced the sectors $Q\neq 0$ which can be regarded as a generalized version of the Ramond sector, and expressed the corresponding bulk theory as $\mathbf{F}$. We also note that the anomaly-free condition and the resultant partition function have been rarely discussed in the literature on category theories.

By studying the anyon condensation \cite{Bais:2008ni,Kong:2013aya,Kawahigashi:2021hds} for each sector $i$ and $a$ in $\mathbf{F}$, we propose the following \emph{bulk semionization} $\{ \Psi\}$ noted as $\mathbf{S}$ corresponding to the SymTFT, 
\begin{equation}
\Psi_{i,p}=\sum_{p'}\frac{\Phi_{i,p'+p,-p'}}{N}\left(=\sum_{p'}\frac{\phi_{i,p'+p}\overline{\phi}_{\overline{i},-p'}}{N}\right) ,
\label{semionization_1}
\end{equation}
\begin{equation}
\Psi_{a,p}=\frac{\Phi_{a,p}}{\sqrt{N}}\left(=\sum_{p'}\frac{\phi_{a,p'+p}\overline{\phi}_{\overline{a},-p'}}{N}\right).
\label{semionization_2}
\end{equation}
As one can see from the above expressions, the summation has been taken corresponding to the addition of the anomaly-free particle $J\overline{J}^{N-1}$, and this can be interpreted as a natural extension of the boson condensation. Whereas the usual anyon condensation results in ring isomorphism or ring homomorphism, the above condensation results in a subalgebra of $\mathbf{F}$. Because of the anomaly-free condition, the total $Z_{N}$ parity and charge are well-defined after the bulk semionization. Hence, one can relate the bosonic or $Z_{N}$ orbifolded modular invariants to the Lagrangian subalgebra of $\mathbf{S}$ \cite{2013PhRvX...3b1009L,Fukusumi:2022xxe,Fukusumi_2022_c} based on them by taking the parity or charge $0$ sector. The constraint from the charge and parity can be regarded as a natural candidate of minimal constraints with a connection to the modular invariants, and this is analogous to the Lagrangian algebra satisfying the maximality condition. More phenomenologically, the condensation should stabilize the system, and these stabilized edge modes are known as protected edge modes. The correspondence between protected edge modes and a modular invariant has been studied in earlier literature \cite{Milovanovic:1996nj,Cappelli:1996np}, and their relation to condensation has been clarified in \cite{2013PhRvX...3b1009L}. One can also see the analogy between this reduction for the modular invariant from the total partition function $\sum_{Q}Z_{Q}$ and dual models in studies on strong interactions\cite{Veneziano:1968yb,Fukusumi:2022xxe,Fukusumi_2022_c}. The modular invariant sector, such as $Z_{0}$, corresponds to the low momentum modes in the dual model, and, by excluding (or gapping out) modular noninvariant parts corresponding to high momentum modes in the dual models, one will obtain a well-organized part of the theory corresponding to the Lagrangian subalgebra. To emphasize the role of subalgebra (or gapping operation) as a variant of anyon condensation, we use Lagrangian subalgebra instead of  ``Lagrangian algebra" in the existing literature.

As from the expressions, one can see that the $Z_{N}$ invariant and noninvariant sectors take totally different form. Moreover, the label of $Z_{N}$ invariant sector $a$ is split or is extended to $(a,p)$ under the $Z_{N}$ extension. Because of this splitting, the theory becomes outside of existing MTCs and nonlocal. For the readers interested in more detailed arguments of this nontrivial aspect, we note the work by the author\cite{Fukusumi:2023psx}.

Here we introduce a deformed bulk theory $\mathbf{F}^{(l)}$ defined by the following set of operators,
\begin{align}
\Phi_{i,p,\overline{p}}^{(l)}&=\phi_{i,p}\overline{\phi}_{\overline{i},\overline{p}}. \\
\begin{split}
\Phi_{a,p}^{(l)}&=\sum_{p'}\frac{\phi_{a,p'+p}\overline{\phi}_{\overline{a},-p'}e^{\frac{2\pi i l \overline{F}}{N}}}{\sqrt{N}} \\
&\left(=\sum_{p'}\frac{\phi_{a,p'+p}\overline{\phi}_{\overline{a},-p'}e^{\frac{-2\pi i l p'}{N}}}{\sqrt{N}}\right).
\end{split}
\label{deformed-Z_n}
\end{align}
where $F$ ($\overline{F}$) is the chiral (antichiral) $Z_{N}$ parity operator of the model\footnote{In this work, we use the notation that the antichiral $Z_{N}$ parity appears in Eq.  \eqref{deformed-Z_n}, but one can freely replace the role of chiral and antichiral part.}, and $l$ is the $Z_{N}$ deformation parameter and takes values $0, 1, ..., N-1$. $\mathbf{F}^{(0)}$ corresponds to the original $Z_{N}$ model. The number $l$ specifies insertion of $Z_{N}$ twisted boundary condition as in \cite{Yao:2020dqx,Fukusumi_2022_c} which is the generalization of $Z_{2}$ case in \cite{Li:2023mmw,Li:2023knf}. We conjecture that this corresponds to the $Z_{N}$ generalization of the even-odd problem in the one-dimensional quantum lattice model\cite{Yang_2004,Hagendorf_2012,Hagendorf:2012fz,Meidinger:2013fqa} (The introduction of \cite{Matsui:2016oqq} contains a concise review of this aspect). By applying the redefinition of the antichiral simple current as $\overline{J}\rightarrow e^{\frac{2\pi i l\overline{F}}{N}}\overline{J}$ for the $Z_{N}$ invariant sector $a$ in the theory $\mathbf{F}^{(l)}$, one can obtain the original theory $\mathbf{F}^{(0)}$. Hence, the existence of the theory $\mathbf{F}^{(l)}$ seems reasonable, but further investigations are necessary. Moreover, it should be stressed that the Hilbert space of $\mathbf{F}^{(l)}$ is different from $\mathbf{F}$ as we demonstrate in Sec. \ref{Vanishing}. One can also obtain the deformed bulk semionization $\mathbf{S}^{(l)}$ as,
\begin{equation}
\Psi_{i,p}^{(l)}=\sum_{p'}\frac{\phi_{i,p'+p}\overline{\phi}_{\overline{i},-p'}}{N}e^{\frac{2\pi il\overline{F}}{N}}=\sum_{p'}\frac{\Phi_{i,p'+p,-p'}}{N} e^{\frac{-2\pi ilp'}{N}},
\label{semionization_3}
\end{equation}
\begin{equation}
\Psi_{a,p}^{(l)}=\sum_{p'}\frac{\phi_{a,p'+p}\overline{\phi}_{\overline{a},-p'}}{N}e^{\frac{2\pi il\overline{F}}{N}}=\frac{\Phi_{a,p}^{(l)}}{\sqrt{N}}.
\label{semionization_4}
\end{equation}

Inversely, one can reconstruct $\mathbf{F}^{(l)}$ from $\{ \mathbf{S}^{(l)}\}$ as,
\begin{align}
\Phi_{i,p,\overline{p}}^{(l)}&=\sum_{l'}\Psi^{(l')}_{i,p+\overline{p}}e^{\frac{-2\pi il' \overline{p}}{N}}, \\
\Phi_{a,p}^{(l)}&=\sqrt{N}\Psi_{a,p}^{(l)}.
\end{align}
This gives a construction of a CFT $\mathbf{F}^{(l)}$ from the sets of TOs $\{ \mathbf{S}^{(l)}\}$, and can be considered as an algebraic version of CFT/TQFT correspondence in \cite{Witten:1988hf,Fuchs:2002cm,Fuchs:2003id,Fuchs:2004dz,Chen:2022wvy}. The total species of the operators in the representation $\mathbf{F}$ and $\mathbf{S}$ is different. To recover a bulk CFT $\mathbf{F}$ in an algebraic way, it is necessary to introduce deformed theories $\{\mathbf{S} ^{(l)}\}$.

\subsection{Lagrangian (sub)algebra}

In this section, we revisit a simple model, the Ising or Majorana CFT, and the corresponding TO in our formalism (A similar argument can be seen in the recent works \cite{Huang:2023pyk,Wen:2024udn,Bhardwaj:2024ydc,Huang:2024ror}, but it contains misleading expressions\footnote{For example, they fail to distinguish the order and disorder fields and use the notations resulting in $1\times 1 \neq 1$}). Compared with most of the other existing works, our algebraic method is totally algorithmic, just the combination of the operations taking the direct product and subalgebra, one can systematically apply the same arguments to the general class of models. Whereas the existing arguments based on category theory or generalized symmetry provide intuitive or phenomenological understanding, it is usually difficult to obtain the algebraic relations, such as the fusion coefficients, without some heuristic arguments.

First, let us introduce the fusion rule of the chiral Ising CFT (or Ising MTC), $\{ I, \psi, \sigma\}$ as follows,
\begin{align}
\psi \times \psi&=I, \\
\psi \times \sigma&= \sigma, \\
\sigma \times \sigma &=I +\psi 
\end{align}
where $I$ is the identity operator in the fusion rule, and the conformal dimensions of the operators are $h_{I}=0, h_{\psi}=1/2, h_{\sigma}=1/16$ respectively. The object $\psi$ is called a chiral Majorana fermion, and $\sigma$ is called a chiral order field in the literature. From these fusion rules, one can read that $\psi$ is the $Z_{2}$ simple current and $\sigma$ is the $Z_{2}$ invariant sector. Hence the sector $\{ I, \psi\}$ corresponds to the label of $i,0$ and $i,1$ and $\{ \sigma \}$ corresponds to the label $a$. The $Z_{2}$ charge of the objects $\{ I, \psi\}$ is zero and that of $\sigma$ is $1/2$. Hence, only the sector $\sigma$ is charged, and this aspect plays a role in the extended theory as we will demonstrate.

By the Moore-Seiberg data, one can obtain the bulk fusion rule as follows,
\begin{align}
\epsilon \times \epsilon&=I, \\
\epsilon \times \sigma_{\text{Bulk}}&= \sigma_{\text{Bulk}}, \\
\sigma_{\text{Bulk}} \times \sigma_{\text{Bulk}} &=I +\epsilon 
\end{align}
where the ring isomorphism is $\{ I, \psi, \sigma\}=\{ I, \epsilon, \sigma_{\text{Bulk}}\}$. These three objects correspond to three sectors of the well-known modular invariant $|\chi_{I}|^{2}+|\chi_{\psi}|^{2}+|\chi_{\sigma}|^{2}$ and correspond to $\{\Phi_{i,p,-p}, \Phi_{a, 0} \}=\mathbf{B}$. $\epsilon$ is called the energy operator and $\sigma_{\text{Bulk}}$ is called the order operator (To distinguish the objects in the bulk CFT and those in the chiral CFT, we have introduced less common notation $\sigma_{\text{Bulk}}$).
The $Z_{2}$ simple current extension introduces the following three objects, which are nontrivial from the bosonic SFC,
\begin{align}
\psi &= \overline{\psi}\times\epsilon,\\
 \overline{\psi} &= \psi \times\epsilon, \\
\mu_{\text{Bulk}} &= \psi \times \sigma_{\text{Bulk}}=\overline{\psi} \times \sigma_{\text{Bulk}}
\end{align} 
By using the relation Eq.\eqref{useful_relation}, one can obtain the fusion rule of the extended model easily. For example, by applying the simple current $\psi$ to the fusion rule of the original SFC $\mathbf{B}$, one can obtain the following relations
\begin{align}
\psi\times (\epsilon \times \epsilon)&=\overline{\psi} \times \epsilon=\psi, \\
\psi \times (\epsilon \times \sigma_{\text{Bulk}})&=\overline{\psi} \times \sigma_{\text{Bulk}}= \mu_{\text{Bulk}}, \\
\psi \times (\sigma_{\text{Bulk}}\times \sigma_{\text{Bulk}})&=\mu_{\text{Bulk}} \times \sigma_{\text{Bulk}} =\psi +\overline{\psi} 
\end{align}

Consequently, the fusion rule $\mathbf{F}^{(0)}$, can be summarized as\cite{Ginsparg:1988ui,Petkova:1988cy,Furlan:1989ra,Runkel:2020zgg},

\begin{align}
\psi\times \psi& =I,\quad \overline{\psi}\times \overline{\psi}=I, \ \epsilon=\psi \overline{\psi}\\
\sigma_{\text{Bulk}}\times \sigma_{\text{Bulk}}&=I+\epsilon, \mu_{\text{Bulk}}\times \mu_{\text{Bulk}}=I+\epsilon, \\
\sigma_{\text{Bulk}}\times \mu_{\text{Bulk}}&=\psi+\overline{\psi}, \\
\psi\times \sigma_{\text{Bulk}}&=\overline{\psi}\times \sigma_{\text{Bulk}}=\mu_{\text{Bulk}}, \\
\psi\times \mu_{\text{Bulk}}&=\overline{\psi}\times \mu_{\text{Bulk}}=\sigma_{\text{Bulk}},
\end{align}
Because of the $Z_{2}$ extension, the three objects in Ising CFT $\{ I, \epsilon, \sigma_{\text{Bulk}}\}$ are mapped to ``$3\times 2=6$" objects, $\{ I, \psi, \overline{\psi}, \epsilon, \sigma_{\text{Bulk}}, \mu_{\text{Bulk}}\}$. One can apply the semionization to this model because the theory is anomaly-free. 

From the fusion rule, one can obtain a subalgebra structure by considering its $Z_{2}$ charge and $Z_{2}$ parity. The $Z_{2}$ charge $0$ sector corresponds to the Neveu-Schwartz (NS) sector of Majorana CFT. Hence, one can obtain the set of operators, $\{I,\psi, \overline{\psi}, \epsilon\}$ with the fermionic partition function $Z_{0}=|\chi_{I}+\chi_{\psi}|^{2}$. By considering the parity and the KW duality, one can obtain the subalgebra $\{I,\epsilon, \sigma_{\text{Bulk}}\}$ or $\{I,\epsilon, \mu_{\text{Bulk}}\}$ corresponding to the bosonic modular invariant $|\chi_{I}|^{2}+|\chi_{\psi}|^{2}+|\chi_{\sigma}|^{2}$. This subalgebra is called Lagrangian (sub)algebra in the contemporary theoretical physics literature\cite{davydov2011structure,2013PhRvX...3b1009L,DavydovMugerNikshychOstrik+2013+135+177,Kaidi:2021gbs,Fukusumi:2022xxe,Fukusumi_2022_c}. However, as can be seen in the Kitaev toric code model\cite{Kitaev:2006lla}, this type of fusion rule cannot appear directly as categorical symmetry\cite{Ji:2019jhk}. 

By applying Eq. \eqref{semionization_1},\eqref{semionization_2}, the generators of the bulk semion algebra of this CFT or SymTFT $\mathbf{S}^{(0)}$ can be constructed as,

\begin{align}
\mathbf{I}&=\frac{I+\epsilon}{2},\ \mathbf{f}=\frac{\psi+\overline{\psi}}{2},\\
\mathbf{e}&= \frac{\sigma_{\text{Bulk}}}{\sqrt{2}},\ \mathbf{m}= \frac{\mu_{\text{Bulk}}}{\sqrt{2}}
\end{align}
where they satisfy the double-semion fusion rule,

\begin{align}
\mathbf{f}\times\mathbf{f}&=\mathbf{I},\\
\mathbf{f}\times\mathbf{e}&=\mathbf{m},\quad \mathbf{f}\times\mathbf{m}= \mathbf{e},\\
\mathbf{e}\times\mathbf{e}&= \mathbf{m}\times\mathbf{m}=\mathbf{I},\ \mathbf{m}\times\mathbf{e}=\mathbf{f}.
\end{align}
In the notation of the general $Z_{N}$ graded SymTFT, the label $\{ \Psi_{i,p} \}$ corresponds to $\{ \mathbf{I}, \mathbf{f} \}$ and the label $\{ \Psi_{a,p}\}$ corresponds to $\{ \mathbf{e}, \mathbf{m} \}$. Because of the extension, the object $\sigma$ in the original Ising MTC splits into the two objects $\{ \mathbf{e}, \mathbf{m} \}$ and the theory becomes outside of the original MTC.

One can straightforwardly check their consistencies by applying the fusion rule of $\mathbf{F}^{(0)}$ and changing the basis, because $\mathbf{S}^{(0)}$ is a subalgebra of $\mathbf{F}^{(0)}$. This simplicity and rigorousness are in sharp contrast with other existing frameworks. For example, one can check the consistency of the new identity operator $\mathbf{I}$ as follows,
\begin{align}
\mathbf{I} \times \mathbf{I}&=\frac{I+\epsilon}{2} \times \frac{I+\epsilon}{2}=\frac{I+\epsilon}{2}= \mathbf{I}\\
\mathbf{I} \times \mathbf{f}&=\frac{I+\epsilon}{2} \times \frac{\psi+\overline{\psi}}{2}=\frac{\psi+\overline{\psi}}{2}= \mathbf{f}\\
\mathbf{I} \times \mathbf{e}&=\frac{I+\epsilon}{2} \times \frac{\sigma_{\text{Bulk}}}{\sqrt{2}}=\frac{\sigma_{\text{Bulk}}}{\sqrt{2}}= \mathbf{e}\\
\mathbf{I} \times \mathbf{m}&=\frac{I+\epsilon}{2} \times \frac{\mu_{\text{Bulk}}}{\sqrt{2}}=\frac{\mu_{\text{Bulk}}}{\sqrt{2}}= \mathbf{m}
\end{align}
We also stress the significance of an algebraic phenomenon that the identity in a subalgebra can be different from the identity of the original algebra. The operator $\mathbf{I}$ is a first example showing this unusual phenomenon as a consequence of condensation in physics. We also note that the coefficients $1/2$ and $1/\sqrt{2}$ are inevitable to obtain the theory $\mathbf{S}$ as a subalgebra of the theory $\mathbf{F}$. For the readers interested in the phenomenological understanding of these coefficients, we note more recent works by the author and collaborators\cite{Fukusumi:2024ejk,Fukusumi:2025clr,Fukusumi:2025ljx}.

By taking the subalgebra $\{ \mathbf{I}, \mathbf{e}\}$ or $\{ \mathbf{I}, \mathbf{m}\}$, one can obtain the electromagnetic duality $\mathbf{e}\leftrightarrow \mathbf{m}$\cite{Huang:2023pyk,Ando:2024hun}. In the same way, one can identify the NS sector as $\{ \mathbf{I}, \mathbf{f}\}$ and Ramond (R) sector as $\{ \mathbf{e}, \mathbf{m}\}$ with T-duality, $\{ \mathbf{I}, \mathbf{f}\}\leftrightarrow \{ \mathbf{e}, \mathbf{m}\}$ implemented by the multiplication of $\mathbf{e}$ or $\mathbf{m}$ to the sectors\cite{Seiberg:2023cdc,Runkel:2020zgg}. This analysis clarifies the relationship between the fusion rule in CFTs and anyons in the TOs or  SymTFT. One can generalize this to $Z_{N}$ models only by considering the $Z_{N}$ parity and charge.

\section{Topological holography and fractional supersymmetry}
\label{TO-CCFT}

In this section, we propose the following correspondence (or ring isomorphism) between the bulk semion algebra and the fusion rule of the CCFT,

\begin{align}
\Psi_{i, p} &\leftrightarrow \phi_{i, p}, \quad (\Psi_{i, p} \leftrightarrow \overline{\phi}_{i, p}) \\ 
\Psi_{a, p} &\leftrightarrow \phi_{a, p}, \quad (\Psi_{a, p} \leftrightarrow \overline{\phi}_{a, p}) 
\end{align}
where we have used the same notations in Sec.\ref{semionization}. This is the algebraic representation of topological holography\cite{Moradi:2022lqp,Bhardwaj:2023bbf,Huang:2023pyk,Inamura:2023ldn,Bhardwaj:2024qrf,Wen:2024udn} relating a $2+1$ dimensional TO denoted by $\{\Psi\}$ and a CCFT denoted by $\{ \phi\}$. As an earlier literature in mathematics, we note \cite{Fuchs:2002cm} with  detailed interpretations of the Moore-Seiberg data\cite{Moore:1988qv,Moore:1989vd}. We also note that the $Z_{N}$ symmetric Cardy states in \cite{Chen:2019sif,Smith:2021luc,Weizmann,Fukusumi:2021zme,Fukusumi_2022_c,Huang:2023pyk} which has a close connection to the bulk topological degeneracies can be labeled by the chiral semions because of the close connection to CCFTs and boundary CFTs (BCFTs)\cite{Saleur:1988zx,Foda:2017vog,PhysRevLett.54.1091,Cardy:1986gw,Lencses:2018paa,Nishioka:2022ook} or the Li-Haldane conjecture in condensed matter\cite{Li_2008,Qi_2012,Cardy:2017ufe}. In existing literature, the extended algebraic structure $\mathbf{S}$ has been obtained heuristically from the algebraic data of $\mathbf{M}$ in some particular cases. Our arguments systematically provide $\mathbf{S}$ as a subalgebra of $\mathbf{F}$, and this is in sharp contrast to the other existing arguments. We stress that even the categorical studies on the extended theory $\mathbf{F}$ are still under development. 

Before moving into the details, we stress again that the exact construction of CCFTs corresponding to TOs has never been achieved in general. This problem can be a central concern in the study of TOs, but there exist theoretical difficulties in CCFTs themselves\cite{Fukusumi:2023psx} because they can be outside of existing MTCs\cite{Lan2016ModularEO,Galindo:2024qzg,Gannon:2024tcl}. However, the bulk CFT is more tractable by the classification of modular invariants\cite{Cappelli:1987xt} and corresponding group extended SFCs \cite{bischoff2021computingfusionrulesspherical,davydov2021braidedpicardgroupsgraded} (and by the conformal bootstrap\cite{Polyakov:1974gs,Belavin:1984vu,Picco:2016ilr,Poland:2018epd}). Our method of constructing $\mathbf{S}$  gives an evident construction of the fusion rule of a CCFT $\{ \phi_{i,p}, \phi_{a,p}\}$ only from the established data of an anomaly-free bulk CFT $\mathbf{B}$ and this gives a first step in studying CCFTs, and the corresponding TOs in a systematic way.

In the next subsection, we demonstrate the benefit of this representation combined with the chiral-antichiral decomposition of a bulk CFT in the Majorana or Ising CFT in a way that can apply to a general class of models. Our method reveals a nontrivial property of (fractional) supersymmetry in the form of a fusion rule with a concrete algebraic expression.

\subsection{Vanishing of fusion rule and (fractional) supersymmetry}
\label{Vanishing}

In the Majorana-Ising CFT, by the chiral (or antichiral) semion $\{ e, m\}$ representation,  one can identify the order and disorder operator in $\mathbf{F}^{(0)}$ as $\sigma_{\text{Bulk}}=(e\overline{e}+m\overline{m})/\sqrt{2}, \mu_{\text{Bulk}}=(m\overline{e}+e\overline{m})/\sqrt{2}$\cite{Ginsparg:1988ui,Fukusumi:2023psx}. Hence, one can reproduce the bulk fusion rule of the original bulk CFT from the combination of the chiral fields, $\{\phi \}=\{ I, \psi, e, m \}=\{ \mathbf{I}, \mathbf{f}, \mathbf{e}, \mathbf{m}\}=\mathbf{S}$, and their antichiral counterparts. We also note that the expression $\sigma_{\text{Bulk}}=\sigma\boxtimes \overline{\sigma}$ can lead to the naive application of the combinations of the fusion and Deligne products resulting in the wrong expression $\psi\times\sigma_{\text{Bulk}}=\psi\times (\sigma\boxtimes \overline{\sigma})=\sigma_{\text{Bulk}}$. It should be stressed that the application of the chiral fermion $\psi$ to the bulk order operator $\sigma_{\text{Bulk}}$ should produce the bulk disorder operator $\mu_{\text{Bulk}}$, and this fact has been established recursively in the studies of both QFT and statistical mechanical models in common. Our expressions reproduce the established result, counting of order and disorder operators, and other existing fusion rules of the Majorana theory in a concrete and rigorous way.

However, there exists another choice of the fusion rule $\mathbf{F}^{(1)}$\cite{Runkel:2020zgg},

\begin{align}
\psi\times \psi& =I,\quad \overline{\psi}\times \overline{\psi}=I, \ \epsilon=\psi \overline{\psi} \\
\sigma'_{\text{Bulk}}\times \sigma'_{\text{Bulk}}&=I-\epsilon, \mu'_{\text{Bulk}}\times \mu'_{\text{Bulk}}=I-\epsilon, \\
\sigma'_{\text{Bulk}}\times \mu'_{\text{Bulk}}&=\psi-\overline{\psi}, \\
\psi\times \sigma'_{\text{Bulk}}&=-\overline{\psi}\times \sigma'_{\text{Bulk}}=\mu'_{\text{Bulk}}, \\
\psi\times \mu'_{\text{Bulk}}&=-\overline{\psi}\times \mu'_{\text{Bulk}}=\sigma'_{\text{Bulk}},
\end{align}
One can reproduce this fusion rule by the identification $\sigma'_{\text{Bulk}}=(e\overline{e}-m\overline{m})/\sqrt{2}, \mu'_{\text{Bulk}}=(m\overline{e}-e\overline{m})/\sqrt{2}$. The corresponding topological defects can be seen in \cite{Li:2023mmw}. We note a consistency check involving the nonabelian fusion rules,
\begin{equation}
\sigma'_{\text{Bulk}} \times \sigma'_{\text{Bulk}}=\frac{e\overline{e}-m\overline{m}}{\sqrt{2}} \times \frac{e\overline{e}-m\overline{m}}{\sqrt{2}}=I -\epsilon
\end{equation}
where we have used the fusion rules of the double semion algebra. One can obtain the other fusion rules of $\mathbf{F}^{(1)}$ in a similar way.

The chiral and antichiral decomposion of $\mathbf{F}^{(0)}$ and $\mathbf{F}^{(1)}$ result in the following (formal) relations,

\begin{align}
\sigma_{\text{Bulk}}\times \sigma'_{\text{Bulk}}&=0,\ \mu_{\text{Bulk}}\times \mu'_{\text{Bulk}}=0, \\
\sigma_{\text{Bulk}}\times \mu'_{\text{Bulk}}&=0,\ \mu_{\text{Bulk}}\times \sigma'_{\text{Bulk}}=0.
\end{align}
This degenerate fusion rule implies that the two representations live in different Hilbert spaces or lattice models. To our knowledge, this degenerate fusion rule has not been studied in literature.

The difference between one-dimensional quantum chains with even and odd sites can be seen in \cite{Yang_2004,Hagendorf_2012,Hagendorf:2012fz,Meidinger:2013fqa} with a close connection to the duality (or supersymmetry) on a lattice model, and this seems to correspond to the above splitting of Hilbert space. Moreover, the operator $\sigma$ and the corresponding topological symmetry can be interpreted as the generator of duality in the model\cite{Frohlich:2004ef,Frohlich:2006ch}. Hence, the following interpretation naturally arises,
\begin{align}
\text{Quantum chain with total cite even}&: \mathbf{F}^{(0)}\\
\text{Quantum chain with total cite odd}&: \mathbf{F}^{(1)}
\end{align}
One can generalize the above arguments to the general $Z_{N}$ or fractional supersymmetric models by studying the $Z_{N}$ parity and charge \cite{Ahn:1990gn,Mohammedi:1994rm,Perez:1996qc,RauschdeTraubenberg:1999tz}. We also note recent works studying similar nontrivial periodicity depending on the size of $Z_{N}$ symmetric lattice models\cite{Watanabe:2021wwt,Hu_2023}.

\section{Relationship between our formalism and existing category theories}
\label{relation_algebra_category}

\begin{figure}[htbp]
\begin{center}
\includegraphics[width=0.5\textwidth]{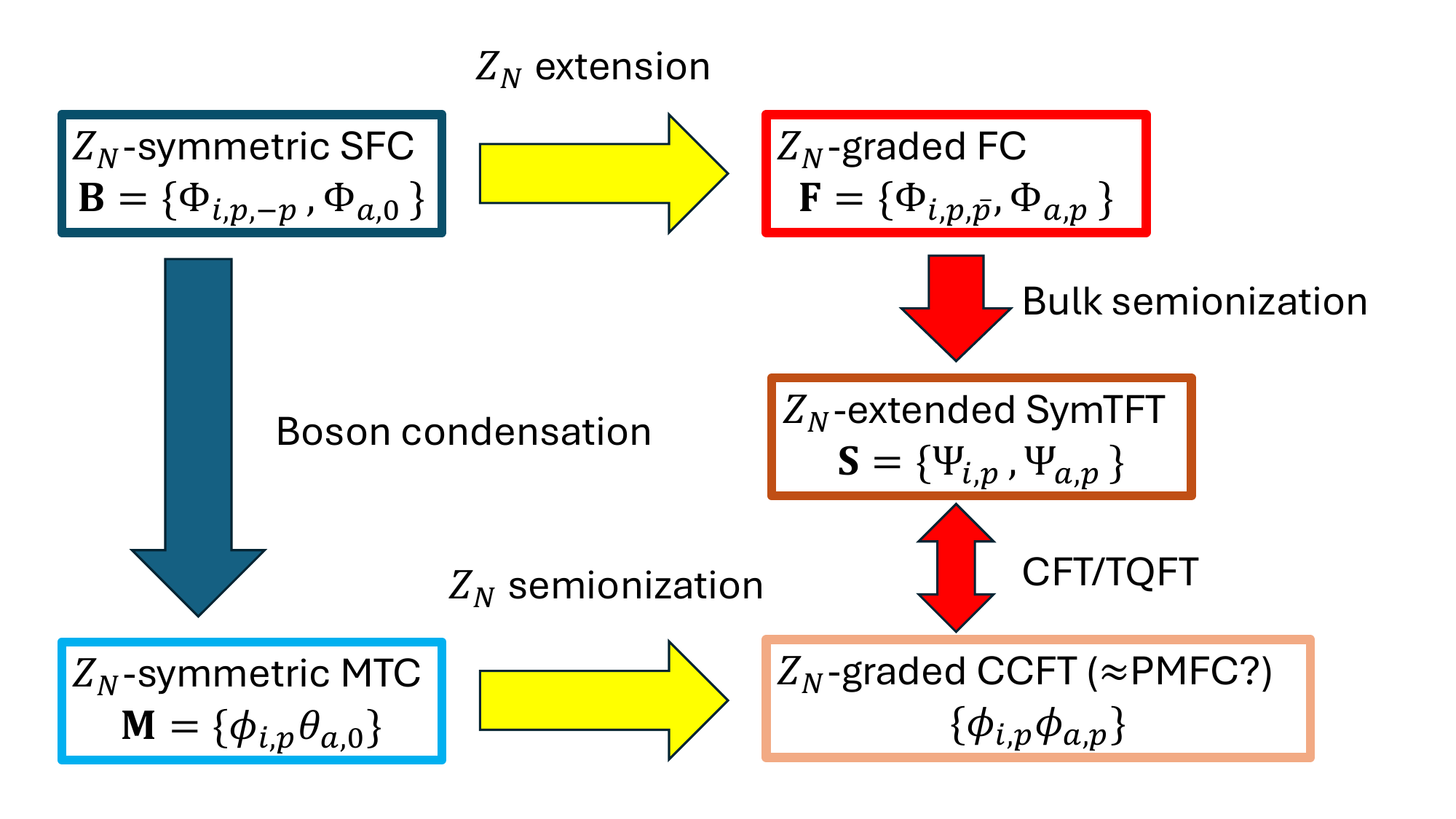}
\caption{Relationship between category theories and our formalism. We note the symbols $\phi, \theta, \Psi, \Phi$ in the corresponding theories to clarify the relations. The blue arrow corresponds to the boson condensation discussed in \cite{Bais:2008ni,Kong:2013aya,Kawahigashi:2021hds}. The yellow arrows correspond to the simple current extension. The respective red arrow or combination of red arrows is called topological holography or sandwich construction in modern literature. The resultant $Z_{N}$ graded CCFT seems to correspond to a premodular fusion category (PMFC)\cite{Wang:2023uoj,Sohal:2024qvq,Ellison:2024svg,Kikuchi:2024ibt}, but further investigations are necessary\cite{Fukusumi:2023psx}. The $Z_{N}$ semionization $\{\theta_{a,0}\}\rightarrow \{\phi_{a,p}\}$ in the figure corresponds to the procedure in the author's works\cite{Fukusumi:2022xxe,Fukusumi_2022_c}  but this is different from $Z_{N}$ extension of MTC in \cite{etingof2009weaklygrouptheoreticalsolvablefusion,Barkeshli:2014cna}. Similar interesting figures can be seen in \cite{Lu:2025gpt,Etingof:2009yvg}, but they are also different from the $Z_{N}$ extension in this manuscript (We thank Zhengdi Sun for the related discussions clarifying this point).
}
\label{category}
\end{center}
\end{figure}

In this section, we clarify the precise relationship between our algebraic formalism and existing category theories. Our construction can be summarized as a construction of a $Z_{N}$ graded fusion algebra (or $Z_{N}$ graded CCFT) $\mathbf{S}$ from $Z_{N}$ symmetric SFC or $Z_{N}$ symmetric bosonic full CFT $\mathbf{B}$. $\mathbf{S}$ provides a new type of of $Z_{N}$ graded algebras corresponding to $Z_{N}$ graded SymTFTs, and they correspond to the modular partition function Eq. \eqref{modular_covariant} which is absent (or at least less common) in literature. We itemize the construction of the $Z_{N}$ extended CCFT in the previous sections using the terminology in category theories as follows.
\begin{itemize}
\item{Choosing spherical fusion category $\mathbf{B}$ with $Z_{N}$ action (or $Z_{N}$ simple current). This corresponds to a bosonic modular invariant and MTC $\mathbf{M}$ \cite{Bais:2008ni,Kawahigashi:2021hds} with a connection to the Verlinde formula\cite{Verlinde:1988sn,Cardy:1989ir}.}
\item{Applying $Z_{N}$ extension to the SFC\cite{turaev2000homotopyfieldtheorydimension,kirillov2001modularcategoriesorbifoldmodels,mueger2004galoisextensionsbraidedtensor,bischoff2021computingfusionrulesspherical,davydov2021braidedpicardgroupsgraded} and obtaining $Z_{N}$ graded fusion category (FC) $\mathbf{F}$. The $Z_{N}$ extension corresponds to the $Z_{N}$ generalization of the parity shift operation in \cite{Runkel:2020zgg}.}
\item{Taking subalgebra of the algebraic data of $Z_{N}$ graded FC and obtaining  $Z_{N}$ extended SymTFT $\mathbf{S}$ as bulk semion algebra. We named this operation bulk semionization.}
\item{Applying CFT/TQFT or topological holography to the resultant bulk semion algebra and obtaining the $Z_{N}$ graded chiral algebra (or CCFT) $\{ \phi\}$.}
\end{itemize}
We summarize the relationship between our formalism and the related category theories in FIG.\ref{category}. The resultant $Z_{N}$ graded chiral algebra $\{ \phi\}$ can be considered as $Z_{N}$ generalization of the superfusion-category \cite{Bruillard:2016yio,bruillard2017classificationsupermodularcategoriesrank,Aasen:2017ubm,Bonderson_2018,Bulmash:2021hmb}, or the ``fractional superfusion category", but further investigations are necessary. (Interestingly, the introduction of $\phi_{a,p}$ results in the emergence of entangled paired particles when considering the chiral correlation functions or corresponding wavefunctions\cite{Stern_2004,Wang:2023uoj,Sohal:2024qvq,Ellison:2024svg,Fukusumi:2023psx}.) We also remind again that the procedure of taking a subalgebra corresponds to the application of the massive RG flow, and this reduces a bulk CFT to (smeared) BCFT\cite{Li_2008,Qi_2012,Cardy:2017ufe}, which corresponds to CCFT under the open-closed duality\cite{Verlinde:1988sn,Cardy:1989ir}. The bulk semionization provides a concrete and concise algebraic representation of this phenomenon, which contrasts with existing analytical methods.

For the readers interested in the status of the theories, we remark that the algebra of anyonic objects usually gives fundamental data of a category theory (and resultant vertex operator algebra). In other words, consistency with anyon algebra is usually required for a category theory corresponding to the models in theoretical physics. In this sense, algebra is more fundamental than category theory in a class of models\cite{Ishikawa:2005ea}, and we list Haagerup CFTs with lattice realization \cite{Huang:2021nvb,Vanhove:2021zop,Corcoran:2024eeh,Bottini:2024eyv} as typical examples (For the  readers interested in the historical aspects, see \cite{Evans:2010yr,Evans:2023nbp} and the references therein). Following these observations, one can state that the correspondence between quantum states, correlation function, and generalized symmetry in bosonic models can be broken in an extended (or gauged) model. In a more recent terminology, these extended models will correspond to symmetry-enriched topological orders.

\section{conclusion}
\label{conclusion}

\begin{figure}[htbp]
\begin{center}
\includegraphics[width=0.5\textwidth]{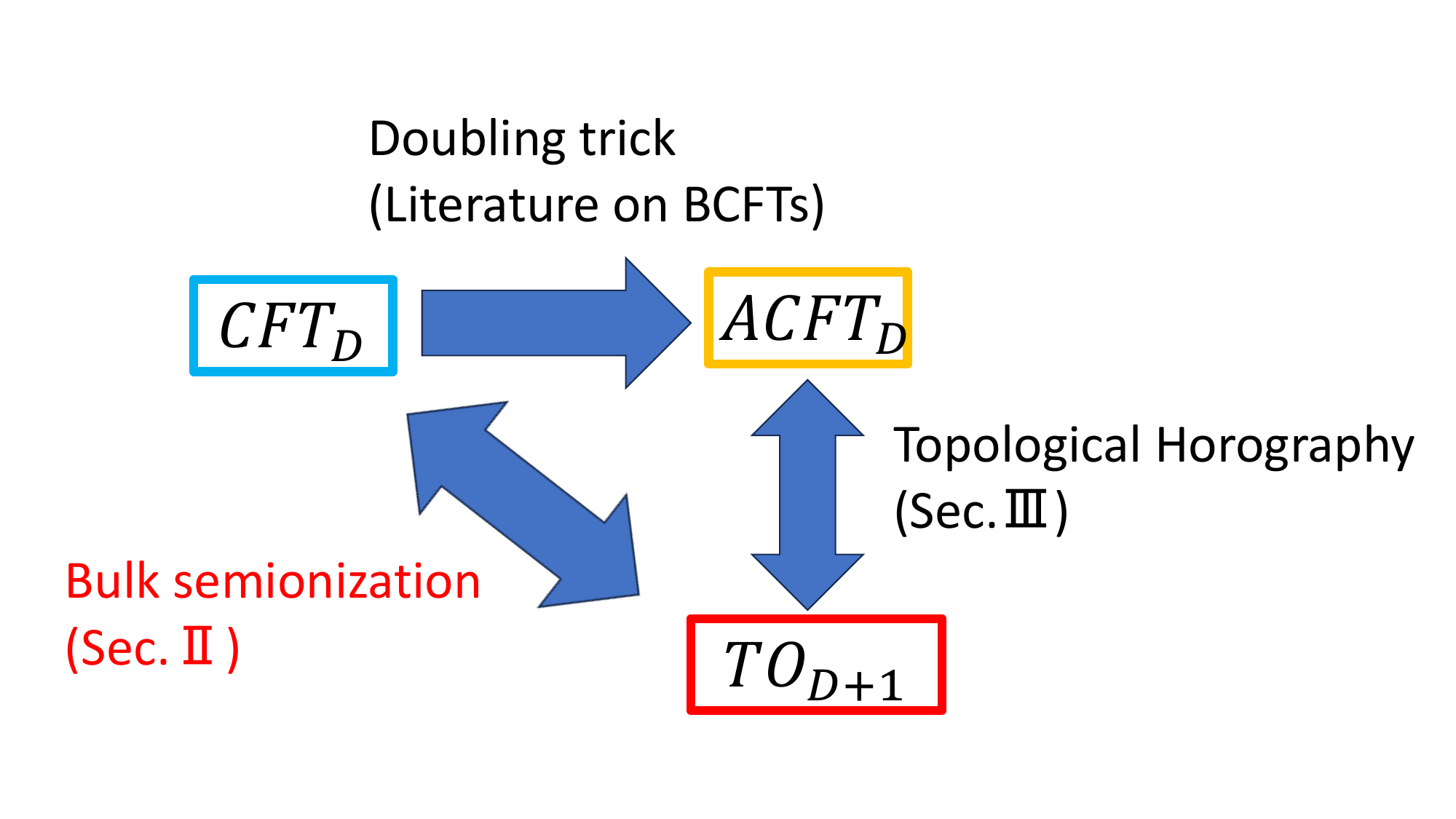}
\caption{A proposal constructing $D$-dimensional Ancillary CFT (ACFT) and $D+1$-dimensional TO from $D$-dimensional CFT. The ACFT proposed in \cite{Nishioka:2022ook} is the generalization of a chiral CFT in general space-time dimensions. The doubling trick \cite{PhysRevLett.54.1091,Cardy:1986gw}, a method to relate a bulk and chiral CFT, has also been revisited in \cite{Nishioka:2022ook}. This is a modification of the similar figure in \cite{Fukusumi:2023psx} by the author. We also note a related proposal by using category theory\cite{Kong:2024ykr}.}
\label{holography}
\end{center}
\end{figure}

In this work, we have studied the structure of fusion rules in CFTs with a modern view and clarified a unified relationship between bulk CFT, CCFT, and TO. Construction of the $Z_{N}$ graded SymTFTs, $\mathbf{S}$, and the analysis of the corresponding partition function \eqref{modular_covariant} are the main results. Especially, the relationship between the bulk and chiral Majorana-Ising CFT and the double semion algebra in the $2+1$-dimensional TO has been established directly by studying the chiral and bulk semionization of the model. Moreover, we have found a new structure, the vanishing fusion rule, and this can explain the duality or (fractional) supersymmetry appearing in contemporary physics in a unified and concise way. Further investigation into this structure on lattice models is an interesting future problem. We expect that the same procedure can apply to  CFTs and the corresponding TOs in general space-time dimensions and modify the recent analysis on topological holography (FIG. \ref{holography}). By combining the Moore-Seiberg data\cite{Moore:1988qv} and symmetry-based analysis of RG domain wall\cite{Gaiotto:2012np} in \cite{Kikuchi:2022ipr}, our method will enable one to unify the ideas in TOs and their gauging structure\footnote{After the submission of the first version of the present work, we noticed interesting applications of CCFT\cite{Fukusumi:2023psx} to the TOs of mixed quantum states\cite{Wang:2023uoj,Sohal:2024qvq,Ellison:2024svg,Kikuchi:2024ibt}. However, it should be noted that their models are still in the scope of the CFT/TQFT combined with the Witt equivalence. This implies the fundamental importance of studying anyons and generalized symmetry in manipulating general mixed quantum states. Related quantum information theoretical aspects which might be useful for this problem can be seen in \cite{Okada:2024qmk}, for example.}. We also note the existence of the exceptional models outside of the Moore-Seiberg data in \cite{Gannon:2001ki,Cappelli:1996np} for further studies.

\section{Acknowledgement}
We thank Takamasa Ando, Hosho Katsura, Ken Kikuchi, Kansei Inamura, Kohki Kawabata, Liang Kong, Simon Lentner, Chihiro Matsui, Yutaka Matsuo, Urei Miura, Sylvain Ribault, Steve Simon, Zhengdi Sun, and Shinichiro Yahagi for the helpful discussions. We especially thank Hosho Katsura for reminding us of the work \cite{Hagendorf_2012}, and Chihiro Matsui for the related lecture course at ``62nd Condensed Matter Physics Summer School (or Bussei Wakate Natsuno Gakkou)", in Japan, 2017. We also thank Guangyue Ji and Bo Yang for the collaboration closely related to this project. We thank many suggestive lectures and discussions in the three independent international conferences, ``Symmetry 2024" in the United Kingdom, ``Conference on Recent Developments in Topological Quantum Field Theory" in China, and ``Focus week on Non-equiribrium physics" in Japan. We thank Jurgen Fuchs, Yuma Furuta, Yasuyuki Kawahigashi, Ingo Runkel, Kareljan Schoutens, Christoph Schweigert, Xiao-Gang Wen, and Masahito Yamazaki for helpful discussions at the conferences. We thank Ingo Runkel and Kareljan Schoutens for notifying us of the literature on anyon condensation and Jurgen Fuchs for the helpful comments on the manuscript. We acknowledge the support from NCTS and CTC.

\appendix

\section{Basis transformations and symmetry in lattice models}
\label{basis}

Recently, non-invertible symmetry in the lattice model has been revisited in \cite{Seiberg:2024gek,Seifnashri:2024dsd,Seiberg:2023cdc}. Their construction seems respective, but can be summarized by using the following chiral (and antichiral) order and disorder fields,
\begin{equation}
\theta_{a,q}=\sum_{p}\frac{\phi_{a,p}e^{\frac{2\pi iqp}{N}}}{\sqrt{N}}
\end{equation}
where $q$ is a kind of bosonic $Z_{N}$ parity, and we used the same notations in the main text. This kind of chiral order or disorder operator has been revisited in the author's related works\cite{Fukusumi:2022xxe,Fukusumi_2022_c,Fukusumi:2023psx}. We also note that this extended bosonic basis and its generalization provide useful representation for the smeared boundary states \cite{Cardy:2017ufe}. (For the readers interested in this aspect, we note more recent works by the author\cite{Fukusumi:2024ejk,Fukusumi:2025ljx}.)

In the Ising CFT, the chiral order operator $\sigma=(e+m)/\sqrt{2}$ (or disorder operator $\mu=(e-m)/\sqrt{2}$) results in the following fusion rules,
\begin{equation}
(\sigma \otimes \overline{\sigma})\times(\sigma \otimes \overline{\sigma})=I+\psi +\overline{\psi}+\epsilon.
\end{equation}
This fusion rule has appeared in \cite{Ellison:2024svg,Seifnashri:2024dsd}, and the argument above shows that the lattice non-invertible symmetry can be reproduced from the field-theoretic argument by representing Rep$({D}_{8})$ as $\{ I, \psi, \overline{\psi},\sigma\otimes \overline{\sigma}\}$. The reason to introduce paired or ``condensed" objects $\sigma \otimes \overline{\sigma}$, $\mu \otimes \overline{\mu}$ (or $\sigma_{\text{Bulk}}$, $\mu_{\text{Bulk}}$) has been clarified in \cite{Fukusumi:2023psx} based on the analysis of operator-state correspondence. The existence of such objects is straightforward by applying the basis transformation of $\sigma \otimes \overline{\sigma}=(\sigma_{\text{Bulk}}+\mu_{\text{Bulk}})/\sqrt{2}$, $\mu \otimes \overline{\mu}=(\sigma_{\text{Bulk}}-\mu_{\text{Bulk}})/\sqrt{2}$. We expect that related lattice objects can be described by the simple current extension of CFTs (or gauged quantum field theories analogous to quark-hadron models) in general.

As in the main text, one can introduce the following objects $\mathbf{L}^{(l)}$which may correspond to the recently proposed symmetry in the lattice models as,
\begin{align}
\Theta_{i,p,\overline{p}}^{(l)}&=\phi_{i,p}\overline{\phi}_{\overline{i},\overline{p}}=\Phi_{i,p,\overline{p}}. \\
\Theta_{a,q}^{(l)}&=\theta_{a,q+l}\overline{\theta}_{\overline{a},-q}=\sum_{p}\frac{\Phi_{a,p}^{(l)}}{\sqrt{N}}e^{\frac{2\pi ipq}{N}}.
\end{align}
In this representation, one can obtain the subalgebra $\{ \Theta_{i,p,\overline{p}}^{(l)},\Theta_{a,0}^{(l)}\}$ for example. This corresponds to a generalization of Rep$(D_{8})$ for the Ising case. We also note that the Rep$(D_{8})$ plays a fundamental role in calculating  correlation functions of the Dirac and Majorana fermion model \cite{Ardonne:2010hj,Zamolodchikov:1987ae,DiFrancesco:1987ez,Fukusumi:2023psx,Hao:2024vlg}.

\section{Gauging and condensation in coupled models: emergence of noninvertible or nonabelian objects}

In this section, we comment on another type of subalgebra related to the gauging procedure\cite{Tachikawa:2017gyf,Bhardwaj:2017xup}. As a simple example, we study  the $SU(2)_{1}\times SU(2)_{1}$ WZW model. The $SU(2)_{1}$ WZW model has two chiral primary fields, $\{ I, j\}$, where $j$ is an anomalous $Z_{2}$ simple current with chiral conformal dimension $1/4$. Hence, by notifying the each copy of $SU(2)_{1}$ by a upper index, the primary fields of $SU(2)_{1}\times SU(2)_{1}$ can be notified as $\{ I, j^{1}, j^{2}, j^{1}j^{2}\}$. Interestingly, one can extract the fusion rule of Majorana-Ising CFT, $\mathbf{F}^{(0)}$, by defining the following object,
\begin{align}
\psi&=j^{1} j^{2}, \ \overline{\psi}=\overline{j^{1}} \ \overline{j^{2}}, \ \epsilon =\psi \overline{\psi}, \\
\sigma_{\text{Bulk}}& =\frac{j^{1}\overline{j^{1}}+j^{2}\overline{j^{2}}}{\sqrt{2}},\ \mu_{\text{Bulk}} =\frac{j^{2}\overline{j^{1}}+j^{1}\overline{j^{2}}}{\sqrt{2}}.
\end{align}
One can obtain the $\mathbf{F}^{(1)}$ by replacing $\sigma'_{\text{Bulk}}=\frac{j^{1}\overline{j^{1}}-j^{2}\overline{j^{2}}}{\sqrt{2}}$ and $\mu'_{\text{Bulk}}=\frac{j^{2}\overline{j^{1}}-j^{1}\overline{j^{2}}}{\sqrt{2}}$.

There exists the related coset representation $SU(2)_{1}\times SU(2)_{1}=\text{Ising}\times SU(2)_{2}$ where $SU(2)_{2}$ has the same fusion rule of that of the Ising CFT\cite{Goddard:1986ee}. In other words, by considering the phase transition or coupled wire construction\cite{PhysRevLett.88.036401,Teo:2011hq} of the Ising CFT or $SU(2)_{2}$ CFT from the $SU(2)_{1}\times SU(2)_{1}$ WZW model, the anomalous $Z_{2}$ symmetry is broken to the noninvertible symmetry $\sigma$. Interestingly, the nonabelian structure of the abelian theory has played a central role in calculating the correlation function of nonabelian theories\cite{Ardonne:2010hj}. As a related research direction, we note that a series of CFTs with the same fusion rule exists when studying Hecke operations or coset representations. For example, one can relate the appearance of the Ising fusion rule to the coset representation $SU(2)_{1}\times SU(2)_{1}=SU(2)_{2}\times \text{Ising}$. 

The same observations can be applied to $\{SU(N)_{1}\}^{k}$ WZW models by decomposing them to Gepner parafermion $SU(N)_{k}/\{U(1)\}^{N-1}$ and $SU(k)_{N}/\{U(1)\}^{k-1}$ and $\{U(1)\}^{N-1}$\cite{Fuji_2017}. This may explain the emergence of noninvertible or nonabelian objects, which can be interpreted as protected edge modes\cite{Graham:2003nc,Fukusumi:2020irh,Fukusumi:2023vjm}. The correspondence between the fusion rules of different CFTs may give clues to understanding the phenomena in TOs and their gauging structures\cite{Tachikawa:2017gyf,Bhardwaj:2017xup}. The conformal embeddings and the corresponding RG domain wall seem to be the most fundamental in this research direction (As one may have already noticed, the argument here is similar to discussions on the Higgs condensation or Nambu-Goldstone boson or fermion. We thank Urei Miura for indicating this view.) For more precise algebraic understanding, we note the recent work by the author\cite{Fukusumi:2025clr}.

\section{Bulk semionization for $SU(3)_{3}$ WZW model}
\label{new_example}

In this section, we apply our method to $SU(3)_{3}$ WZW model with integer spin $Z_{3}$ simple currents. First, we note the data of the chiral fusion rule or MTCs as $\mathbf{M}=\{ \phi_{o,p}, \phi_{j,p}, \phi_{j^{\dagger},p}, \theta_{a}\}$, where the conformal dimensions are summarized as follows:
\begin{align}
Q=0&:h_{o,0}=0, h_{o,1}=h_{o,2}=1, h_{a}=1/2,  \\
Q=1/3&:h_{j,0}=2/9,:h_{j,1}=5/9,h_{j,2}=8/9, \\
Q=2/3&:h_{j^{\dagger},0}=2/9,:h_{j^{\dagger},2}=5/9,h_{j^{\dagger},1}=8/9,
\end{align}
where $Q$ is the $Z_{3}$ charge specifying the sectors. In this model, the $Z_{3}$ simple currents corresponds to $\phi_{o,1}$ and $\phi_{o,2}$, and only the field $\theta_{a}$ is $Z_{3}$ invariant. Corresponding to this structure, one can identify the lower index $p=0,1,2$ as $Z_{N}$ parity. Hence, we focus on the $Z_{3}$ invariant object $\theta_{a}$ and its splitting under the extension and bulk semionization. Some part of the fusion rules with charge cancellation can be seen in \cite{Alimohammadi:1993sy}, and the more general fusion rules are in the scope of the Verlinde formula\cite{Verlinde:1988sn}. Hence, with some patient calculations, one can obtain whole fusion rules. To simplify the discussion, we mainly focus on the fusion rule $\theta_{a}\times \theta_{a}=\sum_{p}\phi_{o,p}+2\theta_{a}$.(In other words, we demonstrate that one can calculate some parts of the extended fusion rules from the given algebraic data without information on the all algebraic data.)

Under the Moore-Seiberg data\cite{Moore:1988qv,Moore:1989vd}, one can obtain the nonchiral fusion ring through the ring isomorphism, $\mathbf{B}=\{ \Phi_{o,p,-p}, \Phi_{j,p,-p}, \Phi_{j^{\dagger},p,-p}, \Phi_{a,0}\}=\{ \phi_{o,p}, \phi_{j,p}, \phi_{j^{\dagger},p}, \theta_{a}\}=\mathbf{M}$.
By applying the $Z_{3}$ extension, the theory can be represented as $\mathbf{F}=Z_{N}\otimes \mathbf{B}=\{ \Phi_{o,p,\overline{p}}, \Phi_{j,p,\overline{p}}, \Phi_{j^{\dagger},p,\overline{p}}, \Phi_{a,p}\}$. Hence, the relation $\theta_{a}\times \theta_{a}=\sum_{p}\phi_{o,p}+2\theta_{a}$ is modified as follows,
\begin{equation} 
\Phi_{a,p}\times \Phi_{a,p'}=\sum_{p''}\Phi_{o,p''+p+p',-p''}+2\Phi_{a,p+p'}
\label{Z_3_F}
\end{equation}
By the bulk semionization, one can introduce the operators in the $Z_{3}$-graded SymTFTs $\mathbf{S}$ as $\Psi_{a,p}=\Phi_{a,p}/\sqrt{3}$, $\Psi_{o, p}=\left(\sum_{p'}\Phi_{o,p'+p,-p'}\right)/3$. Hence, after the bulk semionization $\mathbf{S}\subset \mathbf{F}$, just by deviding the left and right handside of Eq.\eqref{Z_3_F} by $3$, one can obtain the resultant fusion rule in $\mathbf{S}$ as
\begin{equation} 
\Psi_{a,p}\times \Psi_{a,p'}=\Psi_{o,p+p'}+\frac{2}{\sqrt{3}}\Psi_{a,p+p'}.
\end{equation}
Interestingly, the coefficient $2/\sqrt{3}$ inevitably appears. We stress that, if one does not permit the appearance of such noninteger coefficients, the algebraic descriptions of the sandwich construction fail. Moreover, in \cite{Fukusumi:2025clr}, the author demonstrated that even in the $SU(2)_{4}$ WZW model, the simplest model with the $Z_{2}$ bosonic simple current, this kind of unusual fusion coefficient appears. The precise relationship between the unusual fusion coefficients and the boson-fermion statistics may be an interesting problem.

By applying the same method to other fusion rules involving the multiplication of $\theta_{a}$, one can obtain the corresponding fusion rules of the $Z_{3}$-graded SymTFTs. For the other fusion rules only involving $\phi$ of the MTC, the fusion rules are unchanged by the bulk semionization.

\bibliographystyle{ytphys}
\bibliography{fusion}

\end{document}